\newcommand{ \beq }{\begin{equation}}
\newcommand{ \eeq }{\end{equation}}
\documentclass[twocolumn,showpacs,preprintnumbers,amsmath,amssymb]{revtex4}
\usepackage{graphicx}% Include figure files
\usepackage{dcolumn}% Align table columns on decimal point
\usepackage{bm}% bold math

\begin{document}

%\preprint{APS/123-QED}

\title{Vortex dynamics in dilute two-dimensional Josephson junction arrays}% Force line breaks with \\

\author{Md. Ashrafuzzaman, Massimiliano Capezzali  and Hans Beck }
\affiliation{%
Institut de Physique, Universit\'e de Neuch\^atel, Switzerland\\
%*Cellular Biophysics and Biomechanics Laboratory, EPFL, 1015 Lausanne, Switzerland
}%


\begin{abstract}
\noindent
The dynamics of thermally excited vortices in a dilute two-dimensional Josephson junction array where a fraction of the
superconducting islands is missing has been investigated using a multiple trapping model. An expression for the frequency
dependent mobility of vortices has been calculated which allows to obtain the frequency dependent complex electrodynamic
response of the array for different fractions of missing islands.
\end{abstract}


\maketitle
\noindent

\section{Introduction}

Josephson junction arrays (JJA) consist of superconducting islands which are usually arranged on an ordered lattice and
coupled by Josephson interaction. Two-dimensional ($2d$) arrays offer a unique opportunity for studying a
variety of topics in $2d$ physics, such as phase transitions, non-linear dynamics, percolation, frustration and disorder, in
relatively clean experimental realisation. Fabrication of arrays and their basic physical properties have been
described in various articles.$^{1,2}$ The islands  become superconducting at a given transition temperature $T_c^o$. Below
this temperature each island $l$ is characterized by its superconducting wave function
\begin{equation}
\psi_l=\arrowvert\psi_l\arrowvert e^{i\theta_l}
\end{equation}
with its amplitude and its phase. For all matters and purposes one can assume that $\arrowvert\psi_l \arrowvert$ has the same
value in each island, such that the phase is the only relevant variable. The islands are linked to each other by the Josephson
coupling. The potential energy of the array is then given by
\begin{equation}
H=\sum_{\langle ll'\rangle}J_{ll'}[1-\cos(\theta_l-\theta_{l'})]
\end{equation}
The sum can usually be restricted to nearest neighbors in the array and the corresponding Josephson coupling $J$ is related to the critical current $I_c$ by
\begin{equation}
J=\frac{\hbar}{2e}I_c
\end{equation}

A charging energy required when Cooper pairs move from one island to another has to be added to the expression (2). It is given by a
capacitance matrix $C_{ll'}$ coupling the time derivatives of the phases (respectively their conjugate momenta) on different
sites.$^{1,2,3}$ Arrays can then be divided into classical and quantum arrays depending on the ratio of the Josephson coupling
energy to the relevant charging energy. The experimental work we are referring to in this article has been performed on
classical arrays for which charging energies are unimportant. Dynamic phenomena for such arrays are usually described in the
framework of the resistively shunted junction (RSJ) model.$^{2,3,4}$ A resistance matrix $R_{ll'}$ describes the normal currents
flowing between the islands.

Research on disordered JJAs is a particularly exciting field. In an array where some sites are missing the electrical and magnetic
properties of the lattice may change drastically depending on the number of missing sites relative to the whole array. This effect
is related to percolation. About half a century ago this concept was introduced by Flory$^5$ and Stockmayer$^6$ in order to
describe disordered structures.  Later on the geometrical and statistical concepts introduced by Broadbend and
Hammersley$^7$ in connection with the study of diffusion of fluids in random media, introduction of fractal concepts$^8$ and the
development of computers and simulation techniques  as well as theoretical and experimental analysis have contributed to deepening
our understanding of the phenomenon of percolation.

We consider an infinite triangular or square lattice, where each site is randomly occupied by a superconducting island with some probability $p$ and empty with
probability $1-p$. The electrical and magnetic properties of the array depend on the value of this percolation fraction $p$.
At a certain value $p_c$ of $p$ the superconducting properties of the array are destroyed. $p_c$ is called critical
percolation limit. At high concentration of missing sites, the occupied sites are either isolated or form small clusters of nearest-neighbors, and no path
connecting opposite edges of the lattice exists. With increasing $p$, the mean cluster size of the occupied sites increases up to the appearance of an infinite
cluster which connects opposite edges of the whole lattice. The existence of an infinite percolating cluster of occupied sites is connected with the criticality
of the percolation ($p_c$). When $p$ is further increased, more and more sites become part of the infinite cluster of occupied sites, and finally at $p=1$, all
sites belong to the infinite cluster. The percolation threshold depends
on different aspects of the lattice structure like its form (triangular, square or others), dimensionality, and on the type of
percolation (bond, site, etc.). For example, in a triangular $2d$ array, for site percolation, one can analytically$^9$ show that
$p_c= 0.5$, but this value can be different for other dimensions of the array and other types of disorder.

Different types of disorder can be considered:

     a. For ${\bf{bond~disorder}}$ the parameters characterizing the individual junctions, namely the Josphson couplings
       $J_{ll'}$, the junction resistances $R_{ll'}$ and the mutual
       capacities $C_{ll'}$, are random. This type of disorder is realized when the positions of the individual superconducting
       islands deviate in a random way from their ideal lattice positions or if they have random size, since $J_{ll'}$, $R_{ll'}$
       and $C_{ll'}$ depend on the distance between the adjacent superconductors and on their geometry.

      b. For ${\bf{site~ disorder}}$ the properties of the superconducting islands, i.e. their resistance and capacitance to
      ground, are random. More simply, in a  dilute  array certain islands are totally missing.

Dynamic measurements$^{10,11}$ have been done on dilute, or  percolative , arrays for $p$-values somewhat larger than the
percolation threshold $p_c = 0.5$, where the disorder has a marked influence on the properties of the
array.  At temperatures that are low compared to the Berezinskii-Kosterlitz-Thouless (BKT) transition temperature,
$T_{BKT}$ (which in a dilute array is lower than for the regular counterpart, due to the absence of certain bonds, see below)
the equations of motion can be linearized, taking into account only small amplitude excitations. The problem
at hand is then the same as determining the vibrational modes of a disordered harmonic solid, which is a well studied field.$^{12}$
In reference$^{13}$ the impedance $Z(\omega)$ of such a
dilute JJA has been determined by evaluating the dynamic voltage correlation function  by different methods. The coherent
potential approximation (CPA) has proven to be a useful approach for treating dynamic disorder problems.
The inverse of the impedance is expressed by an effective coupling function $K(\omega,p)$ depending on frequency and on the
dilution fraction $p$ :
\begin{equation}
      Z(\omega)=(\eta+\frac{K(\omega,p)}{i\omega})^{-1}
\end{equation}
 corresponding to the effective inductance $L$ and resistance $R$ of the array.

As a main result for the BKT transition these calculations reveal the existence of an effective Josephson coupling constant,
$J_{eff}$. For bonds where one or both of the superconducting islands is missing, the coupling energy vanishes, i.e. $J_{ll'}=0$.
Thermodynamics is then governed by an effective coupling constant $J_{eff}$ given by the relation$^{13}$
\beq
J_{eff}=\frac{p-p_c}{1-p_c}J
\eeq
and thus the transition temperature $T_{BKT}$ ($k_BT_{BKT}=\frac{\pi}{2} J_{eff}$ or $0.9J_{eff}$  for triangular or tetrangular
array respectively) also decreases.

Close to and above $T_{BKT}$ the dynamic response is then dominated by the motion of thermally created vortices.
An equation of motion for vortices can be obtained starting from the RSJ equations for the superconducting phases$^{3}$.
Much analytical and numerical efforts have been devoted$^3$ to describing the dynamic behaviour of the vortices in regular
JJAs. One of the key quantities is the frequency and temperature dependent vortex mobility $\mu(\omega,T)$. For calculating
this quantity the Coulomb interaction between the vortices has to be taken into account, which makes the problem difficult.

For disordered JJAs this is an even more complex problem, since one should now describe the motion of interacting particles in a
random potential landscape. We will treat the vortex response mainly above $T_{BKT}$ where the interaction between vortices
is screened and will thus be neglected. The influence of the random potential on the vortex dynamics will be treated in
the Multiple trapping model, that has been developed for handling the motion of electrons in disordered semiconductors.
This model will be presented in section II. It allows to calculate the average vortex mobility $\mu(\omega, T)$ in the dilute
array as a function of frequency $\omega$ and temperature $T$. In section III the measured electrodynamic response, expressed by
the complex impedance of the array, will be related to $\mu$. In section IV our results for $\mu$ will be presented and
analysed. The main features of our theoretical results for the complex impedance of the array are in good agreement with
experimental data$^{11}$. We shall also show theoretically calculated flux noise spectra for disordered arrays.
Finally in chapter V we shall give some conclusions.

\section{The multiple trapping model}

\subsection{The multiple trapping equations}

In this chapter we discuss our model for the vortex dynamics in dilute $2d$ JJAs (missing sites or bonds). Deviations from a
regular lattice structure will have an influence on the vortices of an array through the Peierls force or pinning potential.
The equilibrium arrangement of the (thermal or field induced) vortices and antivortices will correspond to a minimum of the free
energy in the given random pinning potential landscape. By thermal excitation, in particular above $T_{BKT}$, vortices will
move around in the dilute array thus giving rise to interesting power law behaviour in the dynamic
impedance $Z(\omega)$.$^{14-16}$

Our main goal is therefore to calculate the frequency dependent resistance ($R$) and inductance ($L$) of a dilute array.
In calculating  $R$ and $L$ we have to first find the mobility ($\mu$) of the vortex (V) and antivortex (A) in the disordered
array. For this we use the so called multiple trapping model (MTM), developed for electronic transport in amorphous
semiconductors.$^{17}$ In this model the regions where sites of the array are missing are regrouped into ¥holes of different
shapes and sizes. The motion of a given particle (we will not distinguish between V and A, since  in the absence of
interaction  they undergo the same influence by the random potential) is then described in probabilistic terms. At a given position
$\bf{r}$  in the array one can either be in one of the holes of the array or in a regular region, occupied by superconducting
islands. Thus the state of a particle sitting at $\bf{r}$ is determined by the following probabilities:

$p({\bf{r}},t)\equiv$ probability that a given particle is free at position ${\bf{r}}$ and time $t$, i.e. it is sitting
between holes, in a regular region.

$p_n(t)\equiv$ probability that the particle is trapped in one of the holes which is indexed by $n$.

In the multiple trapping model the time evolution of these probabilities is governed by two rate equations:
\begin{equation}
\frac{\partial p({\bf{r}},t)}{\partial t}+\sum_n\frac{\partial p_n(t)}{\partial t}=-\nabla\cdot{\bf{j}}
\end{equation}
\begin{equation}
with \qquad \frac{\partial p_n(t)}{\partial t}=-\gamma_{r,n}p_n(t)+\gamma_{t,n}p({\bf{r}},t)
\end{equation}

Here $\gamma_{t,n}$ is the probability per unit time (transition rate) for a particle to get trapped inside the hole $n$ and
$\gamma_{r,n}$ is the release rate out of the hole $n$.

If the particle remains in a regular area it contributes to the current ${\bf{j}}={\bf{v}}p({\bf{r}},t)$ with its
velocity ${\bf{v}}=\mu_0{\bf{E}}$ determined by the mobility $\mu_0$ for a regular array of the same structure, and the effective
field ${\bf{E}}$ which will be assumed here as unidirectional (in X direction). Thus the first of the two rate equations can be
written as
\beq
\frac{\partial p(x,t)}{\partial t}+\mu_0E\frac{\partial p(x,t)}{\partial x}=\sum_n\gamma_{r,n}p_n(t)-\sum_n\gamma_{t,n}p(x,t)
\eeq

Now we use Laplace transformation
\begin{equation}
\begin{array}{rcl}
\hat{f}(z)=\int_0^{\infty}dte^{-zt}f(t) \cr  \\
\frac{\partial f(t)}{\partial t}\to z\hat{f}(z)-f(0)
\end{array}
\end{equation}

$f(0)$ being the initial condition.

For (8) this yields
\beq
z\hat{p}(z)-p(0)+\mu_0E\frac{\partial \hat{p}(z)}{\partial x}=\sum_n\gamma_{r,n}\hat{p}_n(z)-\sum_n\gamma_{t,n}\hat{p}(z)
\eeq
and for (7)
\beq
z\hat{p}_n(z)=\gamma_{t,n}\hat{p}(z)-\gamma_{r,n}\hat{p}_n(z)+p_n(0)
\eeq

Imposing the initial condition $p(0)=1$ and $p_n(0)=0$ we get from the previous equation
\beq
\hat{p}_n(z)=\frac{\gamma_{t,n}}{z+\gamma_{r,n}}\hat{p}(z)
\eeq

Now we define
\beq
\pi_n(z)=\frac{\gamma_{t,n}}{z+\gamma_{r,n}}
\eeq

and equation (10) becomes
\beq
z\hat{p}(z)(1+\pi(z))=1-\mu_0E\frac{\partial \hat{p}(z)}{\partial x}
\eeq

The quantity
\beq
\pi(z)=\sum_n\pi_n(z)
\eeq
will turn out to play a central part for the frequency and temperature dependence of the electrodynamic response of the array.

In order to get concrete expressions for our transition rates we consider thermal equilibrium in zero external field where the
probabilities, $p({\bf{r}},t)=p_0$ and $p_n(t)=p_n$, are independent of space and time. The former is simply proportional to the
total regular area $p.(La)^2$, $a$ being the lattice constant, $L^2$ the total number of sites :
\beq
p_0=A.p.(La)^2
\eeq

The probabilities $p_n$ are given by a Boltzmann factor, involving the binding energy $E_n$ of a particle trapped in hole number
$n$, and the surface $S_n$ of the corresponding hole
\beq
p_n=AS_ne^{\beta E_n}
\eeq

The thermally activated release rate is given by
\beq
\gamma_{r,n}\doteq r e^{-\beta E_n}
\eeq
with $r$ being the attempt frequency and $\beta=\frac{1}{k_BT}$.

In the case of detailed balance all terms in equation (7) vanish, therefore
\beq
p_0\gamma_{t,n}=p_n\gamma_{r,n}
\eeq
and $\gamma_{t,n}=\frac{p_n}{p_0}\gamma_{r,n}=r\frac{S_n}{p(La)^2}$.

For the circular holes, considered here, the area is given by $S_n=S(N)=Na^2$, $N$ being the number of missing sites.

Combining these relations we find the following expression for the trapping rate $\gamma_{t,n}=\gamma_t(N)$
\begin{equation}
\gamma_t(N)=r\frac{N}{pL^2}=r\frac{N}{pN_{tot}}
\end{equation}
where $N_{tot}=L^2\equiv$ is the total number of sites.

\subsection{The vortex mobility}

The mean velocity $\bar{v}$ of a V is related to its mean position $\bar{x}$ by
\beq
\bar{v}=\frac{\partial\bar{x}}{\partial t}=\int d^2rx\frac{\partial p({\bf{r}},t)}{\partial t}
\eeq

Through Laplace transformation (9) we get
\beq
\hat{\bar{v}}=z\hat{\bar{x}}=\int d^2rxz\hat{p}({\bf{r}},z)
\eeq
taking, for simplicity,  $\bar{x}(0)=0$, i.e. the vortex started ($t=0$) its motion from the origin of the lattice.
Now using equation (14) the previous equation, through partial integration, becomes
\begin{eqnarray}
\hat{\bar{v}}=\int d^2r x \frac{1-\mu_0E\frac{\partial \hat{p}({\bf{r}},z)}{\partial x}}{1+\pi(z)}&&{}
\nonumber\\
&&\hspace{-4cm}=\frac{-\mu_0E}{1+\pi(z)}\int d^2rx\frac{\partial \hat{p}({\bf{r}},z)}{\partial x}+\int d^2r\frac{x}{1+\pi(z)}{}
\nonumber\\
&&\hspace{-4cm}= \frac{\mu_0E}{1+\pi(z)}
\end{eqnarray}
The last term of the first part in the previous equation represents the initial condition of the velocity and contributes zero
by our assumption.

The effective mobility $\mu$ of a V is given by $\hat{\bar{v}}=\mu E$. We get from the previous equation$^{18}$
\beq
\mu(z)=\frac{\mu_0}{1+ \pi(z)}
\eeq
and $\pi(z)$ is given by equation (15). The purpose of our calculations is to exhibit, through the mobility $\mu$, the
influence of disorder on the dynamic response of the vortex system. In order not to mix these effects with the intrinsic
frequency dependence of $\mu$ in regular arrays  and for computational simplicity  we take $\mu_0$, the vortex mobility in the
regular array, to be constant, giving the scale unit for $\mu$ (see expression (38)).

Taking $D(N)$ to be the number of holes having $N$ missing sites and considering (20) for $\gamma_t(N)$, the key quantity
$\pi(z = i\Omega)$, for real frequencies $\Omega$, can be expressed by
\begin{subequations}
\begin{equation}
\begin{array}{rcl}
\pi(z= i\Omega)\equiv\pi(\Omega)\to\sum_ND(N)\pi(z = i\Omega,N)\cr  \\
=\sum_N\gamma_t(N)\frac{D(N)}{ i\Omega +re^{-\beta E(N)}}\cr  \\
=\int_{1}^{N_{max}}{dN{\hat{D}(N)N\over i{\Omega\over r}+e^{-\beta E(N)}}}
\end{array}
\end{equation}
In the above expression,
$\hat{D}(N)={D(N)\over pN_{tot}}$ is a normalized number density of holes. We introduce the dimensionless frequency
$\omega={\Omega\over\omega_{a}}$,
with $\omega_{a}={k_{B}T_{BKT}\mu_{0}\over a^{2}}$ being a characteristic frequency which
includes the lattice constant a of the array and the bare vortex mobility $\mu_{0}$ of the ordered array. Introducing the vortex diffusion constant by $D_0\equiv k_BT_{BKT}\mu_0$, for temperature $T_{BKT}$,
the frequency $\omega_a$ is given by $\omega_a=\frac{D_0}{a^2}$. Its inverse is thus the average time for a vortex to diffuse
across one lattice constant $a$.
A very sensitive step in our entire procedure is represented by the
choice of the attempt frequency $r$. We will write the latter as $r=\omega_{a}f(t)$ where $t={T\over T_{BKT}}$
represents the scaled temperature.
Here we consider either $f(t)=t$, if the attempt frequency is considered to increase with temperature - i.e.
$r=\omega_{a} t$ -, or
$f(t)=1$, in the case where $r$ is assumed to be constant.
At first glance, the former case probably appears to be more
sound and
intuitive from a physical point of view. However, choosing a constant $r$ may be understood as
arising from the fact that our system is considered to be overdamped; hence,
kinetic energy - which is indeed proportional to temperature -
is an ill-defined quantity and one
may consequently be led to assume, that $r$ does not depend on temperature and simply
represents
a characteristic frequency of the vortex system. We end up with
\begin{equation}
\pi(\omega) =  \int_{1}^{N_{max}}{dN{\hat{D}(N)N\over i{\omega\over f(t)}+e^{-\beta E(N)}}}
\end{equation}
\end{subequations}

The value of $N_{max}$ has to guarantee the correct total number of missing sites :
\beq
N_{tot}(1-p)=\sum_1^{N_{max}} D(N)N
\eeq
Therefore $N_{max}$ depends on the value of $p$, the sample size $N_{tot}$ and of course on the choice of a hole
distribution function $D(N)$.

\subsection{Binding energy for circular holes}

The energy of a phase configuration is expressed by (2). In the case of percolative arrays
\begin{equation}
J_{ll'}=\left\{\begin{array}{ll}
J & \textrm {with probability $p$}\\
0 & \textrm {with probability $1-p$}
\end{array}\right.
\end{equation}
where $J$ is the Josephson coupling constant between two islands.

Now when phase field $\theta(r)$ varies slowly we can write $1-\cos(\theta_l-\theta_{l'})\approx\frac{(\theta_l-\theta_{l'})^2}{2}$,
so we get, converting summation into integration for slowly varying phase fields,
\beq
    E\approx J\int d^2r\frac{(\nabla\theta)^2}{2}
\eeq

For a vortex at the origin, the phase field is given by $\theta(r)=arc\tan\frac{y}{x}$ where $x$ and $y$ are
cartesian coordinates in the array. We get $\arrowvert\nabla\theta\arrowvert^2=\frac{1}{r^2}$ and the total energy of a vortex in the
array is
\beq
E_V=\frac{J}{2}\int_{r_1}^Ld^2r\frac{1}{r^2}
\eeq
where $L$ is the size of array and $r_1$ is a lower cut-off corresponding to lattice constant. Now the binding energy of the V
or A inside a hole (approximated to be circular) of radius $r_2$ with $\pi r_2^2=(N+1)\pi r_1^2$ (the area of the hole is equal
to $N+1$ times  the area of a unit cell) is
\begin{eqnarray}
E(N)&&=\frac{J}{2}\int_{r_1}^Ld^2r\frac{1}{r^2}-\frac{J}{2}\int_{r_2}^Ld^2r\frac{1}{r^2}{}
\nonumber\\
&&{}=\frac{J}{2}\int_{r_1}^Ldrr\frac{1}{r^2}\int_0^{2\pi}d\theta-\frac{J}{2}\int_{r_2}^Ldrr\frac{1}{r^2}\int_0^{2\pi}d\theta{}
\nonumber{}\\&&{}=\pi J\ln\frac{r_2}{r_1}=\pi J\ln\sqrt{N+1}
\end{eqnarray}
because $\pi(r_2^2-r_1^2)=N\pi r_1^2$ or $\frac{r_2}{r_1}=\sqrt{N+1}$.

Using (5)
\beq
 k_BT_{BKT}=\frac{\pi}{2}J_{eff}=\frac{\pi}{2}\frac{p-p_c}{1-p_c}J
\eeq
the exponent of the Boltzmann factor in (25) can be rewritten as
\beq
\beta E(N)=\frac{1-p_c}{p-p_c}\frac{2}{t}\ln\sqrt{N+1}
\eeq

Thus we obtain from equation (25)
\beq
\pi(\omega)=\frac{1}{N_{tot}p}\int_1^{N_{max}}dN\frac{ND(N)}{ i\frac{\omega}{t} +(1+N)^{-a(t)}}
\eeq
with a temperature dependent exponent
\beq
a(t)=\frac{1}{t}\frac{1-p_c}{p-p_c}
\eeq

\section{Electrodynamic response of the array}

\subsection{Resistance and Inductance}

Generally, the response of the array to an external electromagnetic excitation can be characterized taking the contributions of the superfluid, normal electrons and  vortices into account. This corresponds to a two-fluid model, where the medium is described by an inductive superfluid channel in parallel with a dissipative channel.

%%%%%%%%%%%%%%%%%%%%   Figure 1: Equivalent circuit diagram for a 2d superconductor     %%%%%%%%%%%%%%%%%%%%%%%
  \begin{figure} [h]
\let\picnaturalsize=N
\def\picsize{12 cm}
\def\picfilename{RLV.eps}
\ifx\nopictures Y\else{\ifx\epsfloaded Y\else\input epsf \fi
\let\epsfloaded=Y
\centerline{\ifx\picnaturalsize N\epsfxsize \picsize\fi \epsfbox{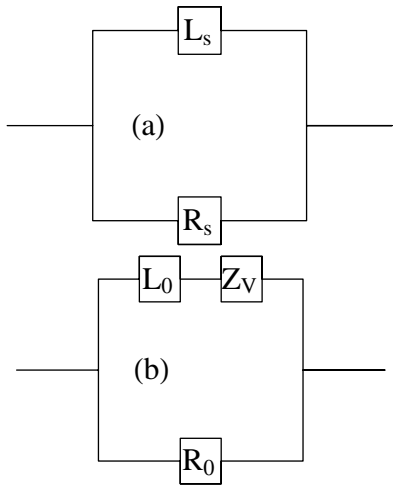}}}\fi
\caption{{\bf Circuit diagram for a $2d$ superconductor. (a) in terms of an inductive and resistive component,
(b) in the presence of vortices.}}
\label{AmplitudePhi4}
\end{figure}
%%%%%%%%%%%%%%%%%%%%%%%%%%%%%%%%%%%   End figure     %%%%%%%%%%%%%%%%%%%%%%%%%%%%%%%%%%%%%%%%

The measured quantity in the array is the sheet conductance $G$. The effect due to the vortices can be incorporated in this
conductance of the array. In the presence of a current, the vortices experience a Lorentz force which will set them in motion
perpendicular to the current flow. Associated with the vortex motion, there is an electric field which adds to the electric field
of the accelerated superfluid background. This phenomenon therefore leads to an increase in the sheet impedance by an amount $Z_V$,
which comes in series with the impedance of the superfluid background, as shown in figures 1(a) and 1(b) schematically.

$G$ is the sheet conductance which is the inverse of the sheet impedance $Z$
\begin{eqnarray}
G &&= \frac{1}{R_s}+\frac{1}{i\Omega L_s}=Z^{-1}{}
\nonumber\\
&&=\frac{1}{R_0}+G_{sup}
\end{eqnarray}
where $G_{sup}$ is the superconducting part of the conductance, the resistance $R_0$ is due to the dissipative processes
(Ohmic contribution of the conductance) resulting from the currents flowing in the junction in the absence of vortices and
antivortices, whereas $L_s$ is the sheet inductance and $R_s$ is the sheet resistance when there are vortices and
antivortices present in the array.

We shall now relate the resistive ($R_s$) and inductive ($L_s$) parts of the sheet conductance $G$ with the resistive ($R_V$) and inductive ($L_V$) parts of the
vortex impedance $Z_V$ through the following relations : $Z_V=Z_{sup}-i\Omega L_0=G_{sup}^{-1}-i\Omega L_0$ and
$G_{sup}=G-\frac{1}{R_0}=\frac{1}{R_s}-\frac{1}{R_0}+\frac{1}{i\Omega L_s}$ so
$Z_V=\frac{1}{R_s^{-1}-R_0^{-1}+(i\Omega L_s)^{-1}}-i\Omega L_0=R_V+i\Omega L_V$ which gives
\beq
R_V=\frac{R_s^{-1}-R_0^{-1}}{(R_s^{-1}-R_0^{-1})^2+(\Omega L_s)^{-2}}
\eeq
\beq
L_V=\frac{1}{\Omega}\frac{(\Omega L_s)^{-1}}{(R_s^{-1}-R_0^{-1})^2+(\Omega L_s)^{-2}}-L_0
\eeq
Here $L_0$, the inductive part, arises from the currents flowing in the junction in absence of vortices and antivortices.

The frequency dependent dielectric function of the vortex system $\epsilon_V(\Omega)=1-iZ_V/(\Omega L_0)$ is related to the vortex
mobility $\mu(\Omega)$ by the following expression
\begin{eqnarray}
 i\Omega L_0\epsilon_V(\Omega)&&=i\Omega L_0(1+2\pi q_0^2 n\mu(\Omega)\frac{1}{i\Omega}){}
\nonumber\\
&&{}=i\Omega L_0+L_02\pi q_0^2n\mu_0(\nu'(\Omega)+i\nu''(\Omega)){}
\nonumber\\
&&{}=i\Omega L_0+Z_V=i\Omega L_0+R_V+i\Omega L_V
\end{eqnarray}
where $q_0 ~(q_0^2=2\pi J)$ is the charge of a vortex, $n$ is the vortex density and
$\nu(\Omega)=\frac{\mu(\Omega)}{\mu_0}=\nu'(\Omega)+i\nu''(\Omega)$ is our dimensionless mobility.

Solving for the real and imaginary parts from the previous equation through the use of equation (5) for the triangular array we get
\beq
R_V=2\pi q_0^2n\mu_0\nu' L_0= 8\pi\omega_a L_0\frac{1-p_c}{p-p_c}\bar{n}\nu'(\omega)
\eeq
\beq
L_V=2\pi q_0^2n\mu_0\frac{\nu''}{\Omega}L_0=8\pi L_0\frac{1-p_c}{p-p_c}\bar{n}\frac{\nu''(\omega)}{\omega}
\eeq
where we have used $\bar{n}=na^2$ which is the density of V per unit cell.

As $Z_V+i\Omega L_0=i\Omega L_0+2\pi q_0^2nL_0\mu_0(\nu'(\Omega)+i\nu''(\Omega))$ we get
\begin{eqnarray}
&&\frac{1}{R_s}-\frac{1}{R_0}+\frac{1}{i\Omega L_s}=\frac{1}{Z_V+i\Omega L_0}{}
\nonumber\\
&&{}=\frac{1}{L_0}\frac{1}{i(\Omega+2\pi q_0^2n\mu_0\nu''(\Omega))+2\pi q_0^2n\mu_0\nu'(\Omega)}{}
\nonumber\\
&&{}=\frac{1}{L_0}\frac{\sigma_V'-i(\Omega+\sigma_V'')}{(\Omega+\sigma_V'')^2+\sigma_V'^2}
\end{eqnarray}
where we have used
\begin{eqnarray}
&&\sigma_V'=2\pi q_0^2n\mu_0\nu'(\Omega)=(2\pi)^2Jn\mu_0\nu'(\Omega){}
\nonumber\\
&&{}=8\pi\omega_a\bar{n}\frac{1-p_c}{p-p_c}\nu'(\Omega)
\end{eqnarray}
\begin{eqnarray}
\sigma_V''=&&2\pi q_0^2n\mu_0\nu''(\Omega)=(2\pi)^2Jn\mu_0\nu''(\Omega){}
\nonumber\\
&&{}=8\pi\omega_a\bar{n}\frac{1-p_c}{p-p_c}\nu''(\Omega)
\end{eqnarray}
where $\sigma_V'$ and $\sigma_V''$ are the real and imaginary parts of  the vortex conductance $\sigma_V$ which is related to vortex dielectric function through
$\epsilon_V(\Omega)=1+\frac{\sigma_V(\Omega)}{i\Omega}$.

Separating the real and imaginary parts from both sides of equation (41) we get$^{18}$
\begin{eqnarray}
&&\frac{1}{R_s}-\frac{1}{R_0}=\frac{\sigma_V'}{L_0((\Omega+\sigma_V'')^2+\sigma_V'^2)}{}
\nonumber\\
&&\hspace{-0.7cm}{}=\frac{8\pi}{L_0 \omega_a}\frac{\frac{1-p_c}{p-p_c}\bar{n}\nu'(\omega)}{(\omega+8\pi\frac{1-p_c}{p-p_c}\bar{n}\nu''(\omega))^2+(8\pi\frac{1-p_c}{p-p_c}\bar{n}\nu'(\omega))^2}
\end{eqnarray}
\begin{eqnarray}
&&\frac{1}{L_s}=\frac{1}{L_0}\frac{\Omega(\Omega+\sigma_V'')}{(\Omega+\sigma_V'')^2+\sigma_V'^2}{}
\nonumber\\
&&\hspace{-1cm}{}=\frac{1}{L_0}\frac{\omega(\omega+8\pi\frac{1-p_c}{p-p_c}\bar{n}\nu''(\omega))}{(\omega+8\pi\frac{1-p_c}{p-p_c}\bar{n}\nu''(\omega))^2+(8\pi\frac{1-p_c}{p-p_c}\bar{n}\nu'(\omega))^2}
\end{eqnarray}

The scale quantity $R_0$ is expressed by
\begin{equation}
R_0=4.5 R_J
\end{equation}
where $R_J$ is the junction resistance and the prefactor 4.5 is an estimate deduced by setting the energy barrier
for vortex motion to its theoretical value (for details see ref.$^{11}$).

The temperature dependent expression for $L_0$, the sheet inductance of a single junction, is given by
(for details see ref.$^{11}$)
\beq
\frac{1}{L_0(T)}=\sqrt{3}\frac{2e}{\hbar}b^{-\zeta}I_c(1-\frac{T}{T_c^o})^2e^{-c\sqrt{T}}
\eeq
% \begin{eqnarray}
% \frac{1}{L_0(T)}&&=\sqrt{3}\frac{2e}{\hbar}b^{-\zeta}I_c(0)(1-\frac{T}{T_c^o})^2e^{-c\sqrt{T}}{}
% \nonumber\\
% &&{}=3.5\times 10^{14} (1-\frac{T}{T_c^o})^2e^{-c\sqrt{T}}
% \end{eqnarray}
Here  $\zeta$ is a critical exponent in two dimensions, $b^{-\zeta}$ is some constant deduced
experimentally by setting the energy barrier for vortex motion to its theoretical value$^{10}$, $I_c$ is the
critical current of a single junction, $c$ is some constant in unit of  $K^{-1/2}$ and  $T_c^o$ is the
transition temperature for the superconducting islands.

\subsection{Flux noise}

Flux noise measurements give interesting information about time correlations in the vortex dynamics. The Fourier transform
$S_{\phi}(\omega)$ of the dynamic correlation function of the magnetic flux threading through a closed loop above the
array is given by$^3$
\begin{eqnarray}
S_{\phi}(\omega)=S_{0}\int_0^{\infty}dk\frac{J_1(kR)^2e^{-2kd}}{k(1+\lambda k)^2}Re\left[\phi_{\rho\rho}(k,-i\omega)\right]
\end{eqnarray}
where $J_1(x)$ is the first order Bessel function and $\lambda$ represents the magnetic penetration depth of the JJA
and $\phi(k,z)$ is the Fourier-Laplace transform of the dynamic correlator of the vortex charge density $\rho_V(r,t)$
\beq
\phi(k,z)=\int d^2r\int_0^{\infty}dt e^{-zt}e^{i{\bf{k}}\cdot{\bf{r}}}\langle \rho_V({\bf{r}},t)\rho_V({\bf{0}},0)\rangle
\eeq
It can be evaluated, for example, by Mori's procedure for calculating dynamic correlation functions, which yields the
following form
\beq
\phi(k,z)=\frac{S(k)}{z+\frac{k_BTk^2\mu(z)}{S(k)}}
\eeq
involving the static charge structure factor $S(k)$ and the dynamic vortex mobility $\mu(z)$. Neglecting again the effect
of vortex interaction we use the mobility resulting from the multiple trapping model for the flux noise calculation.
The structure factor should, in principle, be calculated by taking into account the effect of the ¥random potential
landscape due to the holes on the vortex configuration. We instead take the simple form$^{19}$
\beq
S(k)=\frac{k^2}{k^2+k_0^2}
\eeq
with
\beq
k_0^2=\frac{2\pi q_0^2n}{k_BT}
\eeq
which has the correct behaviour for $k\to 0$ (charge neutrality of the V-A-system) and for $k\to\infty$ .

%%%%%%%%%%%%%%%%%%%%%%%%%%%%%%%%%%%%%%%%%%%%%%%%%%%%%%%%%%%%%%%%%%%%%%%%%%%%%%%%%%%%%%%%%%%%%%%%%%%%
                                             \section{Results}
%%%%%%%%%%%%%%%%%%%%%%%%%%%%%%%%%%%%%%%%%%%%%%%%%%%%%%%%%%%%%%%%%%%%%%%%%%%%%%%%%%%%%%%%%%%%%%%%%%%%

In this chapter we present our theoretical results for the frequency dependent mobility of  vortices or antivortices
in a dilute superconducting array using the multiple trapping model, the conductance of the array and the vortex resistance. Affolter $^{11}$ has measured the properties of such a disordered array near the critical percolation limit $p_c$
below which the superconductivity of the array destroys (a publication of his results is in preparation).

In order to compare our theoretical results with these experimental data we have to determine the values of the parameters characterizing the array.

The measurements $^{11}$ have been performed on a triangular JJA  of lead islands at $p=0.515$ with the following characteristics: $R_J\approx 5.2m\Omega$, $\zeta\approx 1$, $a\approx 15\times 10^{-6}m$, $I_c(0)\approx 14 mA$, $c\approx 3 K^{1/2}$,
$T_c^o=7K$, $T_{BKT}=3.7K$, $b\approx 0.22$. From this we find : $D_0 \approx  2\times 10^{-5}\frac{m^2}{s}$ and $\omega_a\approx 10^{6}$Hz.

%In reference$^{11}$ a different reduced temperature $\tau=k_BT/J(T)$ is used, based on an effective, temperature dependent Josephson coupling $J$.
%The link between $\tau$ and $T$ is approximately given by $\tau=e^{1.8(T-6.04)}$. This relation allows to evaluate the quantity $L_0(T)$ given by
%expression (47). The link between $\tau$ and $t$ is given by the fact that at $T = T_{BKT}$ we have: $t = 1$ and $\tau_c = 0.055$
%(the relation follows from $T=6.04+\frac{1}{1.8}\ln(0.055t)$).
%This information will be needed in order to compare our theoretical results with the measured data in the following figures.

An important goal consists in elucidating the way in which the disorder of the dilute array influences vortex motion.
The key quantity for this is $\pi(\omega)$ given by expression (33). Given the limits of the
integral over $N$ three frequency regimes can be distinguished, namely
\begin{equation}
\begin{array}{rcl}
(a)~\omega<\omega_1\equiv (1+N_{max})^{-a(t)} \cr  \\
(b)~\omega>\omega_2\equiv 2^{-a(t)} \cr  \\
(c)~\omega_1<\omega<\omega_2
\end{array}
\end{equation}

In regions $a$ and $b$ the frequency dependence of the real and imaginary parts, $\pi'(\omega)$ and $\pi''(\omega)$,
does not depend on the details of the hole distribution $D(N)$. It is given by
\begin{equation}
\begin{array}{rcl}
Region~a:\quad \pi'(\omega)=\pi_0,\quad \pi''(\omega)\propto\omega \cr  \\
Region~b:\quad \pi'(\omega)\propto\frac{1}{\omega^2},\quad \pi''(\omega)\propto\frac{1}{\omega}
\end{array}
\end{equation}

In the intermediate region, however, the form of $D(N)$ will be reflected in temperature dependent frequency
exponents for  $\pi'(\omega)$ and $\pi''(\omega)$. A simple estimate can be made by assuming a power law
distribution of hole sizes:
\beq
D(N)=D_0N^{-s}
\eeq
which turns out to correspond quantitatively with the arrays, studied in Ref.$^{11}$, where $s\approx 1.8$.
For frequencies lying well inside the interval [$\omega_1,\omega_2$] the power law of $\pi$ can be simply calculated:
\beq
\pi(\omega)\propto\int_1^{N_{max}}dN\frac{ND(N)}{ i\omega +(1+N)^{-a(t)}}\approx I_0(i\omega)^{-u(t)}
\eeq
\beq
I_0=\int_0^{\infty}dy\frac{y^{1-s}}{1+y^{-a}}
\eeq
\beq
u=1+\frac{2-s}{a}
\eeq

%%%%%%%%%%%%%%%%%%%%%%%%%%%%%%%%%%%%%%%%%%%%%%%%%%
\begin{table}[h]
\def\picsize{12 cm}
\begin{tabular}{|c|c|c|c|c|c|}
\hline
$t$ & $a(t)$ & $\omega_1$ & $\omega_2$ & $u(t)(s=1.8)$ & $u(t)(s=0.3)$\\
\hline
0.9 & 37 & $10^{-53}$ & $7.10^{-12}$ & 1 & 1.05\\
2 & 17 & $9.10^{-25}$ & $8.10^{-6}$ & 1.01 & 1.1\\
6 & 5.5 & $2.10^{-8}$ & 0.02 & 1.04 & 1.3\\
10 & 3.3 & $2.10^{-5}$ & 0.1 & 1.06 & 1.5\\
\hline
\end{tabular}
\caption{{Some parameters for $p=0.515$ and $N_{max}=25$; $a(t)=\frac{1}{t}\frac{1-p_c}{p-p_c}=\frac{1}{t}\frac{0.5}{0.015}=\frac{33}{t}$,
$\omega_1=(1+25)^{-a}$ and $\omega_2=(1+1)^{-a}$.  }}
\end{table}

%{$t$ & $a(t)$ & $\omega_1$ & $\omega_2$ & $u(t)(s=1.8)$ & $u(t)(s=0.3)$}

We summarize in Table I the values of $\omega_1,~\omega_2$  and $u$ using the parameters corresponding to the
experimental data of Ref.11, namely $p=0.515,~0.9<t<10$ and $N_{max} = 25$. For the given parameter values
it turns out that the exponent $u\approx 1$
is almost $T$-independent. The resulting frequency dependence the real part of $\pi$, $\pi'\sim1/\omega$,
is unusual, since one expects $\pi'$ to be an even function of $\omega$, as it is indeed the case in regions
$a$ and $b$, see relations (54). The power law behaviour ($\pi\sim\omega^{-u}$) is the result of the integration
over a contiuous spectrum of relaxation rates in (56). For comparison we give another form of $D(N)$
which gives more weight to large holes ($s=0.3$). The corresponding exponent has a more pronounced $T$-dependence.

%%%%%%%%%%%%%%%%%%%%%%%%%%%%%%%%%%%%%%%%%%%%%%%%%%%%%%%%    Figure 5: Expt. Rs    %%%%%%%%%%%%%%%%%%%%%%%%%%%%%%%%%%%%%%%%%%%
%	\begin{figure} [h]
%	\let\picnaturalsize=N
%	\def\picsize{7 cm}
%	\def\picfilename{Rsexpt.eps}
%	%\def\picfilename{Rss18p053.eps}
%	\ifx\nopictures Y\else{\ifx\epsfloaded Y\else\input epsf \fi
%	\let\epsfloaded=Y
%	\centerline{\ifx\picnaturalsize N\epsfxsize \picsize\fi \epsfbox{\picfilename}}}\fi
%	\caption{{\bf $R_s$ vs $\omega$ plot for $p=0.515$ taken from experimental data.$^{11}$ $s=1.8$. $\tau$ is the reduced temperature and $\tau_c=0.055$ is the critical reduced temperature.}}
%	\label{AmplitudePhi4}
%	\end{figure}

The vortex mobility resulting from $\pi$ through (24) will determine the electrodynamic response in (44) and (45),
as well as flux noise (48). Its behaviour, in units of the free vortex mobility $\mu_0$, is shown in
figure 2 for different temperatures and for different defect concentrations $1-p$. The three frequency regimes are
again visible, although the curves have more structure than the ones for $\pi$ since $\pi'(\omega)$ and
$\pi''(\omega)$ are combined with each other in the expression (24) for $\nu'(\omega)$ and $\nu''(\omega)$.

%%%%%%%%%%%%%%%%%%%%%%%%%%%%%%%%%%%%%%%%%%%    Fig2:  mobility   %%%%%%%%%%%%%%%%%%%%%%%%%%%%%%%%%%%%%%%%%
\begin{figure} [h]
\let\picnaturalsize=N
\def\picsize{30 cm}
\def\picfilename{mobs18p0515po6DispapFig2.eps}
\ifx\nopictures Y\else{\ifx\epsfloaded Y\else\input epsf \fi
\let\epsfloaded=Y
\centerline{\ifx\picnaturalsize N\epsfxsize \picsize\fi \epsfbox{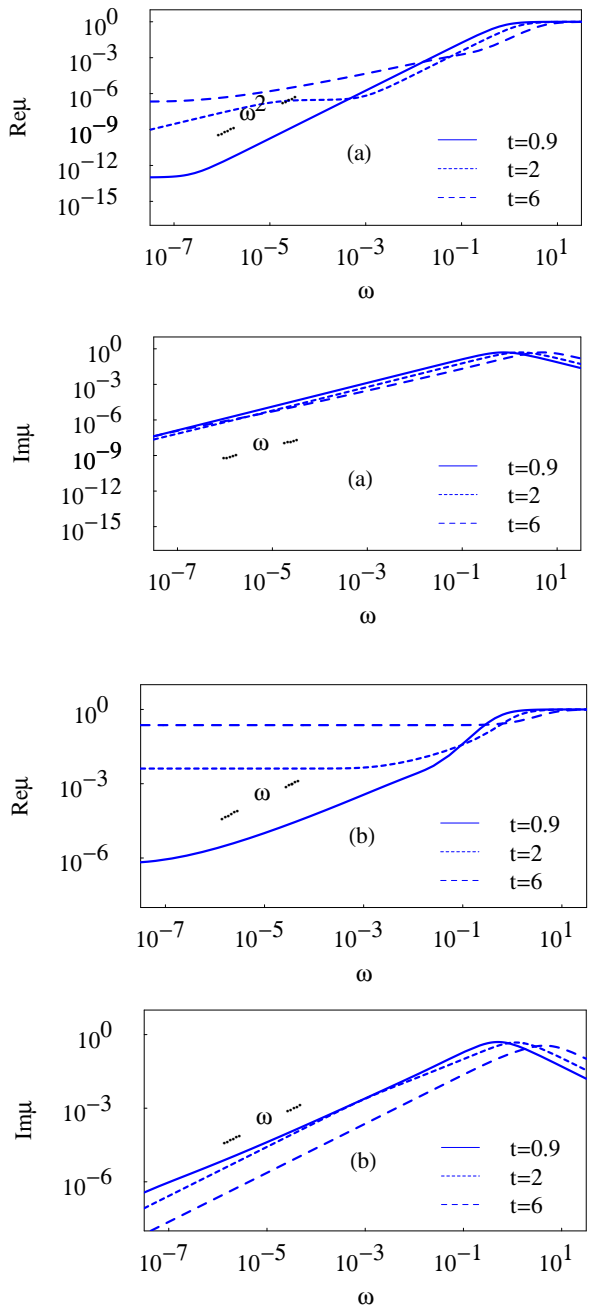}}}\fi
\vspace{-0cm}
\caption{{\bf $\mu$ vs $\omega$  for (a) $p=0.515$ ($N_{max}=25$) and for (b) $p=0.6$ ($N_{max}=22$). $s=1.8$. In order
to see the frequency dependence easily the slopes for $\omega$, respectively $\omega^2$, are given. }}
\label{AmplitudePhi4}
\end{figure}

At very high frequencies, namely $\omega>1$ the real part of the mobility is constant with the bare value $\mu_{0}$.
In this regime we are investigating the short-time response of the array : the vortices lying
outside the holes do not have a significant probability of being trapped and their mobility is just the bare one. At the other extremity of the frequency spectrum, the vortex mobility is also constant but the bare value $\mu_{0}$ is renormalized by $1+\pi_{0}$ leading to a reduction by several orders of magnitudes, strongly depending on $p$. Over very long time spans the vortices get trapped and released many times and, as they do not contribute to mobility as long as they stay in a hole, the overall mobility dramatically diminishes upon increasing the number and size of the holes.

The imaginary part of the vortex mobility displays a maximum near $\omega = 1$. For lower frequencies $Im[\mu(\omega)]\propto\omega$, although at $p=0.6$ and below $T_{BKT}$ a small kink appears around $\omega\sim 10^{-7}$ (especially for $s=0.3$).

The intermediate-frequency behaviour of the real part of the vortex mobility can be understood by looking
carefully at the interplay of the real and imaginary parts of $\pi(\omega)$ going
into $\mu'(\omega)={1+\pi'(\omega)\over (1+\pi'(\omega)^{2})+\pi''(\omega)^{2}}$.

Two frequency domains show up : In a range below $\omega = 1$, extending as far down as $10^{-7}$ for $t=0.9$,
$\pi''(\omega)\propto\omega^{-1}$ for all interesting frequencies, while $\pi'(\omega)$ becomes significantly
lower than unity. Thus the real part of the mobility increases quadratically with frequency, which is
in some sense  a  surprisingly normal behaviour. Conversely, at smaller frequencies, $\pi'(\omega)$
becomes constant, while $\pi''(\omega)\propto\omega$ (i.e. it reaches a maximum at some frequency, below the
quadratic-behaviour window), giving rise to a constant vortex mobility with a value intermediate between $\mu_0$ and
the low frequency limit. The two frequency regimes get shifted towards lower $\omega$-values when temperature
is reduced, as it can be seen in figure 2. The curves in figure 2 have been calculated for the choice
$f(t)=t$ in equation (25b). We have verified that the result for $f(t)=1$ is almost identical.

In addition to the intermediate-frequency regime, where the real part of the mobility becomes constant,
we see on Fig. 2, that at $t=2$ yet another anomalous regime shows up at very low frequencies, where
$\mu'(\omega)\propto\omega$, approximately. At high temperature ($t=6$) and at frequencies below the tiny $\mu\propto\omega^2$
window centered around $\omega\approx 1$, the mobility plateau gets replaced by a regime where $\mu'(\omega)\propto\omega^{0.8}$,
down to $\omega\approx 10^{-6}$. The non-integer exponent arises from the fact that, in accordance with the
Table I, $\pi''(\omega)\propto\omega^{-1.04}$ in this frequency window, while $\pi'(\omega)\propto\omega^{-1}$
and $\pi'(\omega)>>1$ still. Obviously, this non-integer power-law behaviour of the real part of the mobility
also arises at lower temperatures, but it becomes less pronounced as the exponent $u(T)$ is extremely close to unity
 at $t=2$. In order to confirm this behaviour, we have computed the vortex mobility at $s=0.3$, a choice for which the exponent
 $u(T)$ has a much more marked temperature-dependence than at $s=1.8$. see Table I. In the former case, we find
 that, at $t=0.9$, no relevant changes occur, with respect to the results obtained at $s=1.8$. Nevertheless, at higher
temperature $t=2$, the intermediate-frequency plateau discussed above gets substantially reduced and covers
at most only one frequency decade. At even higher temperature $t=6$, the plateau is replaced by an anomalous regime
extending roughly from $\omega\approx 10^{-7}$ to $\omega\approx 10^{-2}$, in which the real part of the mobility
behaves as a power-law $\omega\propto\omega^{-0.7}$ again with a non integer exponent. For both choices of the parameter $s$,
the real part of the mobility eventually turns to a constant at extremely low frequencies. Therefore, in conclusion, for
temperatures above $T_{BKT}$, the mobility displays four distinct frequency regimes, which collapse into three
regimes below the transition temperature.

%%%%%%%%%%%%%%%%%%%%%%%%%%%%%%%%%%%%%%%%%%%%%%%%%%%%%%%%    Figure 3: Rs    %%%%%%%%%%%%%%%%%%%%%%%%%%%%%%%%%%%%%%%%%%%
  \begin{figure} [h]
\let\picnaturalsize=N
\def\picsize{30 cm}
\def\picfilename{Rss03p0515DispapFig3.eps}
\ifx\nopictures Y\else{\ifx\epsfloaded Y\else\input epsf \fi
\let\epsfloaded=Y
\centerline{\ifx\picnaturalsize N\epsfxsize \picsize\fi \epsfbox{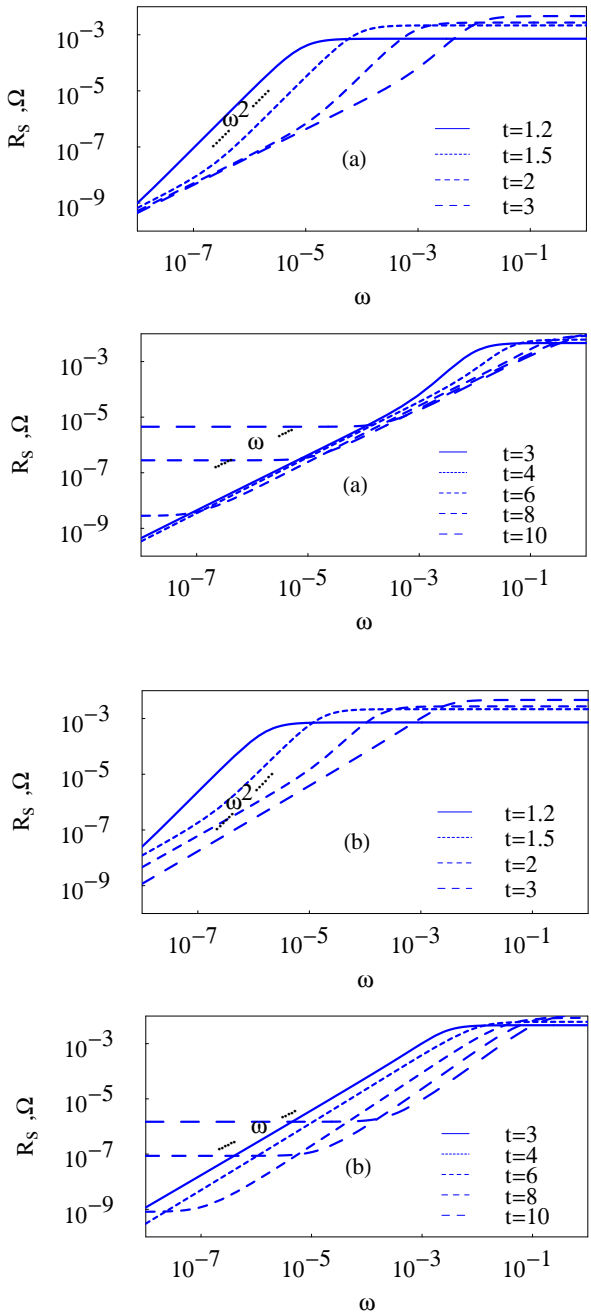}}}\fi
\vspace{-0cm}
\caption{{\bf $R_s$ vs $\omega$ for $p=0.515$. (a) $s=1.8$ and (b) $s=0.3$.}}
\label{AmplitudePhi4}
\end{figure}

%%%%%%%%%%%%%%%%%%%%%%%%%%%%%%%%%%%%%%%%%%%%%%%%%%%%%%%%    Figure 4: Rs    %%%%%%%%%%%%%%%%%%%%%%%%%%%%%%%%%%%%%%%%%%%
  \begin{figure} [h]
\let\picnaturalsize=N
\def\picsize{17 cm}
\def\picfilename{Rss18p053.eps}
\ifx\nopictures Y\else{\ifx\epsfloaded Y\else\input epsf \fi
\let\epsfloaded=Y
\centerline{\ifx\picnaturalsize N\epsfxsize \picsize\fi \epsfbox{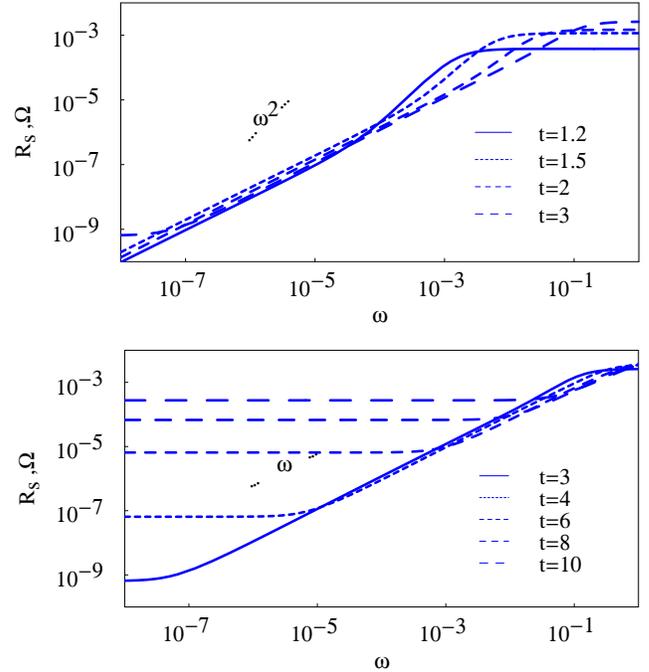}}}\fi
\vspace{-1cm}
\caption{{\bf $R_s$ vs $\omega$ plot for $p=0.53$ ($N_{max}=25$) and $s=1.8$.}}
\label{AmplitudePhi4}
\end{figure}

We now turn to the electrodynamic response for which experimental data are available in ref.$^{11}$. For the vortex density
$n$ showing up in (44) and (45) we have used the values obtained  in the references$^{20,21}$ through Monte Carlo
simulations of the regular arrays. In order to have a consistent treatment one should, of course, know the vortex density
of a dilute array. The presence of holes indeed makes the formation of phase singularities more easy, since, in particular,
vortices centered in a hole cost less energy than in a regular array.  However, for temperatures above $T_{BKT}$, where
even in a regular array the number of vortices grows rather rapidly, the difference should not be too important.
Our sheet resistance curves for $p=0.515$ are shown in figures 3(a) and 3(b) for two different exponents $s$ in
$D(N)$, see equation (55). For the sake of comparison figure 4 shows the same curves for a slightly larger value
of $p$. The following observations can be made:

- The same three frequency regimes determining the quantity $\pi(\omega)$ and the vortex mobility can be identified in $R_s(\omega)$ : at very low and at very high frequencies $R_s$ is constant, whereas in the intermediate regime it
increases as a power of $\omega$: $R_s\sim\omega^{x(T)}$. This temperature dependent exponent $x$ is a fine
detail of the MT model which is in good agreement with the experimental data. For higher temperature $x=1$,
which is again an interesting signature of the disorder of the array. For lower $T$ the sheet resistance
undergoes an upturn to $\omega^{2}$-behaviour, which is clearly visible in Fig. 3. The latter behaviour is intimately linked to the real part of the vortex mobility (discussed above) going through an intermediate plateau in some frequency window.
Indeed, within the latter, we notice that $\mu_{V}'(\omega)\ll\mu_{V}''(\omega)$, trivially implying $\sigma_{V}'(\omega)\ll\sigma_{V}''(\omega)$.
Upon considering Eqns. (42) and (43) for $p$ close to $p_{c}$, we furthermore deduce that $\sigma_{V}''(\omega)$ is larger
than frequency in the "plateau" $\omega$-window. Consequently, inserting this into expression (44), we obtain
$R_{s}^{-1}={L_{0}\sigma_{V}''(\omega)^{2}+R_{0}\sigma_{V}'(\omega)\over R_{0}L_{0}\sigma_{V}''(\omega)^{2}}$. Hence,
bearing in mind that for all the frequencies that we are considering
here, $\mu_{V}''(\omega)\propto\omega$, there exists a frequency window in
which $\mu_{V}'(\omega)=const.$ implies $R_{s}\propto\omega^{2}$,
especially for temperatures close to and above $T_{BKT}$, where $L_{0}(T)$ is
small. In turn, at higher frequencies, we fall in the $\omega$-window where $\mu_{V}'(\omega)\propto\omega^{2}$, in which the sheet resistance turns to a constant. In conclusion, the anomalous vortex mobility plateau which arises from the balance between the real and imaginary parts of $\pi(\omega)$ is at the root of the $\omega^{2}$-upturn of the sheet resistance in our model.

%%%%%%%%%%%%%%%%%%%%%%%%%%%%%%%%%%%%%%%%%%%    Fig6:  Ls  %%%%%%%%%%%%%%%%%%%%%%%%%%%%%%%%%%%%%%%%%
\begin{figure} [h]
\let\picnaturalsize=N
\def\picsize{17 cm}
\def\picfilename{Ls18s03p0515.eps}
\ifx\nopictures Y\else{\ifx\epsfloaded Y\else\input epsf \fi
\let\epsfloaded=Y
\centerline{\ifx\picnaturalsize N\epsfxsize \picsize\fi \epsfbox{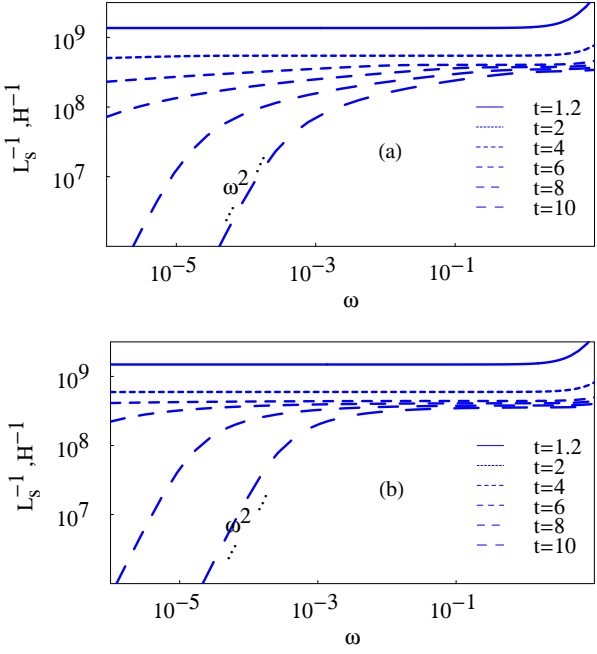}}}\fi
\vspace{-1cm}
\caption{{\bf $L_s$ vs $\omega$ plot for $p=0.515$. (a) $s=1.8$, (b) $s=0.3$.}}
\label{AmplitudePhi4}
\end{figure}

- In experiment $R_s$ approaches a square root frequency dependence for $\omega$ larger than $\omega_c$ (see ref.$^{11}$),
rather than becoming constant. This may point to a new dynamic regime that is not fully covered by our model calculations. There is however a growing tendency in the
Theoretical curves to such a further increlase of $R_s$ at the highest frequencies shown for larger p-values (see figure 4).

- The temperature variation of the slow frequency level of $R_s$, as well as the difference between the low and high
frequency limits are larger in the MTM results than in experiment. This may be a hint that the model,
at least for the $p$- value used, attributes too much value to the disorder of the array.

- For higher temperatures the measured data show universality: when the curves have reached  a linear slope  they lie
on top of each other. In the model results this universality is rather well reproduced for a higher value of
the exponent $s$ of the hole distribution function $D(N)$ (figures 3(a) and 4), which corresponds to the
effective hole hierarchy of the experimental array$^{11}$, giving thus less weight to large holes than the lower
values of $s$.

%%%%%%%%%%%%%%%%%%%%%%%%%%%%%%%%%%%%%%%%%%%    Fig7:  Ls   %%%%%%%%%%%%%%%%%%%%%%%%%%%%%%%%%%%%%%%%%
\begin{figure} [h]
\let\picnaturalsize=N
\def\picsize{9 cm}
\def\picfilename{Ls18p53.eps}
\ifx\nopictures Y\else{\ifx\epsfloaded Y\else\input epsf \fi
\let\epsfloaded=Y
\centerline{\ifx\picnaturalsize N\epsfxsize \picsize\fi \epsfbox{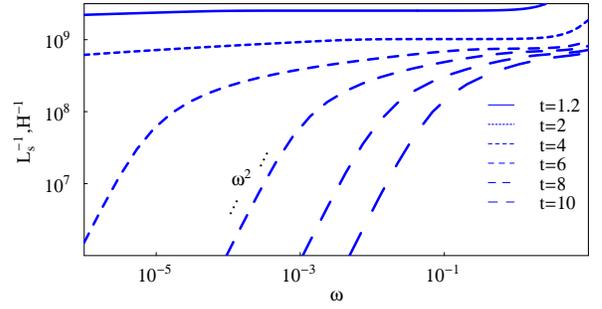}}}\fi
\vspace{-0cm}
\caption{{\bf $L_s$ vs $\omega$ plot for $p=0.53$. $s=1.8$.}}
\label{AmplitudePhi4}
\end{figure}

The inverse of the inductive response, $1/L_s$,  for our MTM is presented in figures 5(a) and 5(b) for the same
parameters as in figure 3 for $R_s$ and for a large $p$ in figure 6. Comparing the two leads to the
following observations:

- The three frequency regimes are again visible: for small and for large $\omega$ $1/L_s$ varies as $\omega^2$,
whereas inbetween it is constant. The experimental data$^{11}$ show this behaviour for low frequency. However,
except for the lowest temperature, there is no real constant part, but relatively smooth cross-over to a square root like
behaviour when $\omega$ increases. This again points to a high frequency dynamics that is not covered by our MT
model for $p=0.515$. However, the experimentally obtained $\sqrt{\omega}$ type behaviour is almost achieved in our theoretical model
for higher values of $p$ and $T$.

%%%%%%%%%%%%%%%%%%%%%%%%%%%%%%%%%%%%%%%%%%%    Fig9: Rv(T)    %%%%%%%%%%%%%%%%%%%%%%%%%%%%%%%%%%%%%%%%%
\begin{figure} [h]
\let\picnaturalsize=N
\def\picsize{17 cm}
\def\picfilename{Rvs18p0515p053DispapFig7.eps}
\ifx\nopictures Y\else{\ifx\epsfloaded Y\else\input epsf \fi
\let\epsfloaded=Y
\centerline{\ifx\picnaturalsize N\epsfxsize \picsize\fi \epsfbox{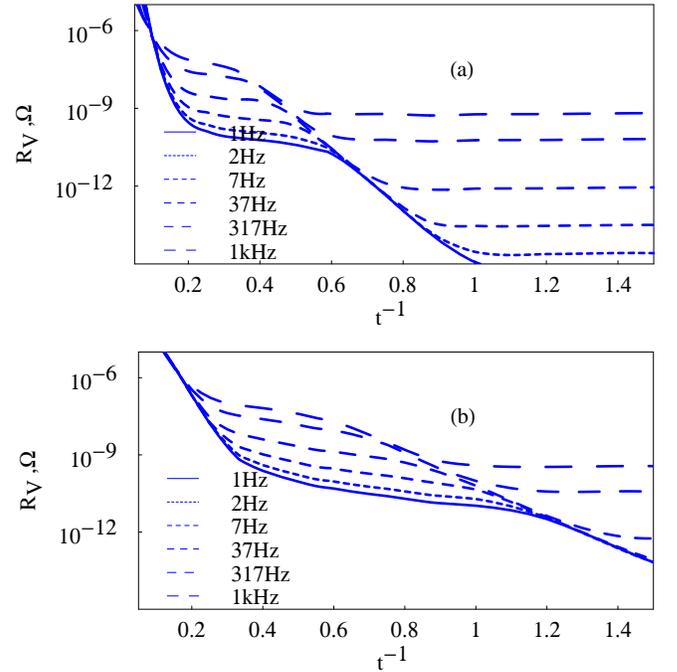}}}\fi
\vspace{-1cm}
\caption{{\bf $R_V$ vs $1/t$  for different $\Omega$. (a) for $p=0.515$, (b) for $p=0.53$. $s=1.8$.}}
\label{AmplitudePhi4}
\end{figure}

On the whole our MTM results are in good agreement with the measured data. Certain experimental features (in particular the low
frequency and low temperature behaviour of $R_s$) are better reproduced for the value $p=0.515$ of the percolation parameter,
which corresponds to the experimental array. Other details of the data are closer to the theoretical curves for
$p=0.53$. This may be due to the oversimplification of the true, ramified structure of the regions of missing sites,
which in the model are approximated by circular hole. On the other hand, concerning the probability distribution $D(N)$
of these holes the exponent $s=1.8$, corresponding to the experimental situation, is more satisfactory than a form for $D$
which would give more weight to large holes.

In section IIIA, equations (36) and (37), we have introduced the notion of a vortex impedance
$Z_V=R_V+i\Omega L_V$. At low frequencies its real part is thermally activated as shown in figure 7. This
behaviour is confirmed by the measurements presented in ref.$^{11}$. The slope of the theoretical curves at high enough
temperature (e.g., around $t=5$) yields
activation energies of about 4 and 2.6 in unit of $J$ for $p=0.515$ and $p=0.53$ respectively, in
good agreement with the experimental data. For very low temperatures the curves become flat. However,
in between the two limiting regimes there is another common slope in the $R_V$ vs $1/t$ plot at about
$1/t=0.6$ for $p=0.515$ and $1/t=1.1$ for $p=0.53$ in our theoretical plots.
The measurements do not really exhibit our second common slope but the curves in ref.$^{11}$ for 317 Hz and
lower frequencies do indeed show a tendency to a steeper slope before they approach the constant value in
the low temperature limit. In this respect the experimental data look more similar to our curves for a
larger percolation fraction ($p = 0.53$ in figure 7).

%%%%%%%%%%%%%%%%%%%%%%%%%%%%%%%%%%%%%%%%%%%    Fig13:  flux noise   %%%%%%%%%%%%%%%%%%%%%%%%%%%%%%%%%%%%%%%%%
\begin{figure} [h]
\let\picnaturalsize=N
\def\picsize{16 cm}
\def\picfilename{fluxp0515p06s18.eps}
\ifx\nopictures Y\else{\ifx\epsfloaded Y\else\input epsf \fi
\let\epsfloaded=Y
\centerline{\ifx\picnaturalsize N\epsfxsize \picsize\fi \epsfbox{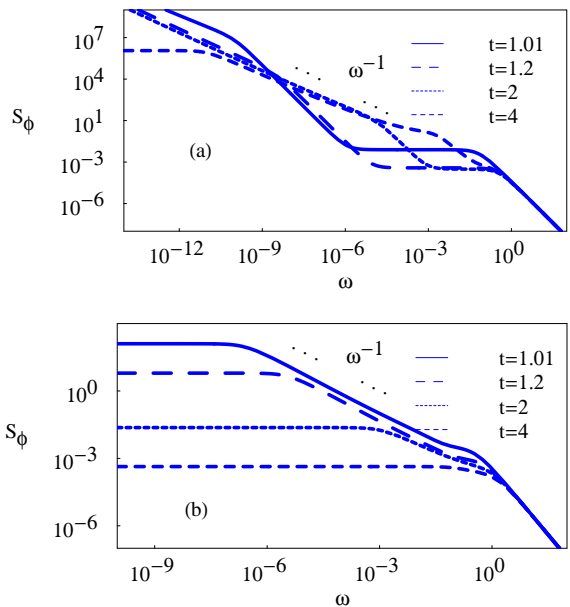}}}\fi
\vspace{-1cm}
\caption{{\bf Flux noise $S_{\phi}$ vs $\omega$ for (a) $p=0.515$ and (b) $p=0.6$. $s=1.8$.}}
\label{AmplitudePhi4}
\end{figure}
%%%%%%%%%%%%%%%%%%%%%%%%%%%%%%%%%%%%%%%%%%%%%%%%%   End Fig   %%%%%%%%%%%%%%%%%%%%%%%%%%%%%%%%%%%%%%%%%%%%%%%%%%%%%%%

Finally we present, in figure 8, the flux noise spectrum resulting from our multiple trapping approach, for different temperatures
and $p$. Flux noise is an interesting observable even for regular arrays$^{3,19}$.
It is white (frequency independent) for sufficiently low $\omega$ with a value that depends on $T$. For
intermediate frequencies the curves for different $T$ are identical, having  a $1/\omega$ slope. This
¥universality has been explained in different ways, for example by invoking the dynamic response of bound
vortex-antivortex pairs which exist below $T_{BKT}$,$^{19}$ but also above, with a finite life-time. It is interesting to note that
disorder seems to produce a similar behaviour, although the intermediate regime has structure superposed on a simple
$1/\omega$ dependence, specially for $p$ close to $p_c$. In particular, there is another white noise region for
higher frequency, before the behaviour crosses over to $1/\omega^2$ at high frequencies. This structure is again a consequenc
of the frequency dependence of the vortex mobility.  Moreover, the universality in this regime is not perfect: even
for $p=0.6$ the curves are not on top of each other. Unfortunately there are, at least for the time being, no experimental data to compare with.

\section{Summary and conclusions}

We have aimed at explaining the electrodynamic response of a Josephson junction array in which a substantial fraction
of the superconducting sites is missing. For the motion of thermally excited vortices and antivortices in this type of strongly
disordered array we have used a multiple trapping model. The regions of missing sites are grouped into holes
which we take to be of circular shape. When a vortex or antivortex reaches a hole it has some probability of
being trapped into the hole (pinning effect). It can later on be released due to thermal excitation.
The effect of this disorder is phrased in terms of a frequency dependent vortex mobility which determines observable quantities, such as the electrical conductance,
composed by the sheet resistance $R_s$ and the inductance $L_s$ of the array, or the flux noise.
We find the following results :

(i) $R_s$ exhibits three frequency regimes. At very low and high frequency $\omega$ a white spectrum is
observed with a much lower value for low $\omega$ than for large $\omega$. For the time scale corresponding to the latter
regime vortices remain either trapped in a hole of free outside any hole.  Thus the mobility, and the
corresponding response is given by the one of a regular array. At the opposite end, for very low frequencies,
vortices get trapped many times during one excitation cycle, which strongly reduces their mobility.
For intermediate frequencies $R_s\propto \omega^{x}$ with $x$ varying between $x=1$ and $x=2$.
For higher temperatures the $R_s\propto\omega$ almost coincide, signalling some kind of universality. These features are in good agreement with the experimental data in $^{11}$.

 (ii) For the inductive part we find $L_s^{-1}\propto\omega^2$ at low $\omega$ while for intermediate frequency
 regime $L_s^{-1}$ is independent of $\omega$. For large frequencies a tendency of $L_s^{-1}$ to grow with frequency
 with some new exponent is also seen for higher values of $p$. The critical frequency for crossover from $L_s^{-1}\propto\omega^2$ to constant
$L_s^{-1}$ decreases with the decrease of temperature $T$ and  percolation fraction $p$. These results also  reproduce well the
experimental data of reference$^{11}$.

(iii) For a given frequency the vortex resistance is thermally activated at sufficiently high temperatures.
The activition energy on the order of 4 times the Josephson coupling between the existing superconducting
sites corresponds to the measured value. For lower temperatures a tendency towards another activated form of
the $R_V$ versus $1/T$ curves appears, which is also visible as a tendency in the measured data.

(iv) We have also evaluated the frequency dependent flux noise.
$1/\omega$ noise is achieved for the intermediate frequencies separating the white noise part for small
$\omega$ and $1/\omega^2$ noise for very high frequencies. thus the presence of disorder leads to similar results
as seen in regular arrays.$^{19}$. However at very close to $p_c$ the flux noise data has some unexpected flatenning
before turing to $1/\omega^2$ part which is a consequence of the frequency dependence of the vortex mobility.
Unfortunately there are no experimental flux noise data for the disordered arrays to compare with our
theoretical investigations.

{\bf{Acknowledgments}}

We thank Piero Martinoli and Philippe Curty for interesting discussions. This work was supported by the Swiss National
Science Foundation under the project no. 2000-067853.02/1.

{\bf{References}}
%\hline

\noindent
$^1$R. S. Newrock, C.J. Lobb, U. Geigenm$\ddot{u}$ller, M. Octavio,

Solid State Physics 54, 263 (2000).\\
$^2$P.Martinoli. Ch. Leemann, Journal of Low

Temperature Physics 118, 699 (2000).\\
$^3$Md. Ashrafuzzaman and H. Beck, Studies of High

Temperature Superconductors (Nova Science,

New York, 2002), Vol.43.\\
$^4$H. Beck and D. Ariosa, Solid state communications

80, 657 (1991).\\
$^5$P.J. Flory, J. Am. Chem. Soc. 63, 3083 (1941).\\
$^6$W.H. Stockmayer, J. Chem. Phys. 11, 45 (1943).\\
$^7$S. R. Broadbend and J. M. Hammersley, Proc. Camb.

Phil. Soc. 53, 629 (1957).\\
$^8$B.B. Mandelbrot, The fractal geometry of nature,

Freeman, San Francisco (1977).\\
$^9$A. Bunde and S. Havlin, Fractal and Disorder Systems,

Springer-Verlag, Berlin (1991).\\
$^{10}$A.L. Eichenberger, J. Affolter, M. Willemin, M.

Mombelli, H. Beck, P. Martinoli, S.E. Korshunov;

Phys Rev Lett 77, 3905 (1996).\\
$^{11}$J$\acute{e}$r$\hat{o}$me Affolter, PhD thesis, Institute of Physics,

University of Neuchatel, Switzerland (2001),

and to be published.\\
$^{12}$T. Nakayama, K. Yakubo, R.L. Orbach ; Rev Mod

Phys 66, 381 (1994).\\
$^{13}$M. Mombelli, H. Beck ; Phys Rev B57, 14'397 (1998).\\
$^{14}$M.P.A. Fisher, Phys Rev Lett 62, 1415 (1989).\\
$^{15}$M.V. Feigel'man, V.B. Geshkenbein, A.I. Larkin,

V.M. Vinokur; Phys Rev Lett 63, 2303.\\
$^{16}$M. Calame, S.E.Korshunov, Ch.Leemann, P.Martinoli;

Phys Rev Lett  86, 3630 (2001).\\
$^{17}$H. Scher and E.W. Montroll, Phys Rev B12, 2455

(1975).\\
$^{18}$Md. Ashrafuzzaman, H. Beck, Journal of

Magnetism and Magnetic Materials (JMMM),

Vol 272-276P1 pp 284-285.\\
$^{19}$Md. Ashrafuzzaman, M. Capezzali and H. Beck,

Phys Rev B 68, 052502 (2003).\\
$^{20}$A. Jonsson and P. Minnhagen, Phys Rev B55, 14

(1997).\\
$^{21}$ Md. Ashrafuzzaman, Ph. Curty, M. Neef and H. Beck,

Inst. of Physics, University of Neuchatel,

Switzerland (2002), unpublished.

%$^{15}$Claude Bailat and Hans Beck, Institute of Physics, University of Neuchatel, Switzerland (1994), Unpublished.

% \hspace{0.6cm}

\end{document}